\documentclass[fleqn,usenatbib]{mnras}

\usepackage{newtxtext,newtxmath}

\usepackage[T1]{fontenc}

\DeclareRobustCommand{\VAN}[3]{#2}
\let\VANthebibliography\thebibliography
\def\thebibliography{\DeclareRobustCommand{\VAN}[3]{##3}\VANthebibliography}

\usepackage{graphicx}	
\usepackage{amsmath}	

\title[Mass of Mira~B]{Is Mira~B a low mass white dwarf? }
\author[Zamanov et al.]{
R. K. Zamanov,$^{1}$\thanks{E-mail: rzamanov@nao-rozhen.org} 
V. Irincheva,$^{1}$
B. Spassov,$^{1}$
A. Kurtenkov,$^{1,2}$
D. Marchev,$^{3}$
M. Minev,$^{1}$ 
M. F. Bode,$^{4,5}$
\newauthor
   L. Dankova,$^{1,2}$
B. Borisov,$^{3}$
M. Dechev,$^{1}$
and  G. Yordanova$^{3}$
\\
\\
$^{1}$Institute of Astronomy and National Astronomical Observatory, 
                     Bulgarian Academy of Sciences, 
                     72 Tsarigradsko Shose, 1784 Sofia, Bulgaria \\
$^{2}$Faculty of Physics, Sofia University "St. Kliment Ohridski", 
             5 James Bourchier Blvd., 1164 Sofia, Bulgaria \\
$^{3}$Department of Physics and Astronomy, Shumen University "Episkop Konstantin Preslavski", 115 Universitetska Str., 9700 Shumen, Bulgaria \\
$^{4}$Astrophysics Research Institute, Liverpool John Moores University, 
                 IC2, 149 Brownlow Hill, Liverpool, L3 5RF, UK \\
$^{5}$Office of the Vice Chancellor, 
      Botswana International University of Science and Technology, 
      Private Bag 16, Palapye, Botswana
}

\date{Accepted .... 2026. Received 8 January 2026; in original form ....}

\pubyear{2026}

\begin{document}
\label{firstpage}
\pagerange{\pageref{firstpage}--\pageref{lastpage}}
\maketitle

\begin{abstract}
We report  photometric observations in  Johnson  $B$ and $V$ bands of the short term 
variability (flickering) of Mira (omicron Ceti). 
The observations were performed during  
7 nights in the period August - October 2025, 
in the course of the last minimum of the Mira pulsations.
The observed peak-to-peak amplitude of the flickering is 0.11-0.28 mag 
in $B$ band.
For the flickering source we find luminosity in the range $0.10-0.46~L_\odot$. 
Using the amplitude-flux relation, 
we estimate an average luminosity of the accretion disc $L_d = 0.91 \pm 0.28 $~$L_\odot$. 

Assuming that the white dwarf accretes material 
through Wind Roche Lobe Overflow (WRLOF),  we find that  
Mira~B is a low mass white dwarf with $M_{wd} = 0.24 \pm 0.04$~$M_\odot$
accreting at a rate $6.8 \times 10^{-9}$~$M_\odot$~yr$^{-1}$. 
This value of the mass is in the range of the extremely low mass white dwarfs. 

The data are available on Zenodo: zenodo.org/records/18756532
\end{abstract}



\section{Introduction}
\label{s.int}
Mira (omicron Ceti, HD~14386) 
is a binary system consisting of an asymptotic giant branch 
star of spectral type  M5-9III (Mira A) 
and a hot companion (Mira~B).
"The Wonderful Star" (omicron Ceti)
was identified as a variable star in 1596 \citep{1933JRASC..27...75H}.
The companion (Mira~B) was discovered in 1922 by  A. H. Joy and  R. G. Aitken 
\citep{1923PASP...35..323A} as a blue star 
located at an angular distance of $\approx 0.6$ arcsec. 
An bow shock and turbulent wake extend over $2^\circ$ on the sky,
arising from Mira's large space velocity
and the interaction between its strong wind and the interstellar medium.
This wind wake is a tracer of the past 30,000 years of Mira's mass-loss
\citep{2007Natur.448..780M}. 
Another part of the wind is captured by Mira~B and 
forms a bridge between Mira~A and Mira~B
\citep{2006ESASP.604..183K}.  

In the recent catalogues, 
Mira is classified  as an accreting-only symbiotic star 
\citep{2019ApJS..240...21A,  2019AN....340..598M},  
in other words it consists of a red giant star and a white dwarf companion which accretes matter from the primary component. 
In the older catalogues a classification as "symbiotic-like" or "weakly symbiotic" can also be found 
\citep[e.g.][]{2000A&AS..146..407B}. 
The accretion produces rapid light-variations on a time scale minutes-hours \citep{1957IAUS....3...46W, 1972MNRAS.159...95W}. 
The amplitude of these optical brightness fluctuations 
is the same as those from accreting white dwarfs in Cataclysmic Variables, 
and significantly larger than one would expect from an accreting main sequence star
and reveals Mira~B to be a white dwarf  \citep{2010ApJ...723.1188S}.

Numerical models \citep{2017MNRAS.468.3408D} and high resolution observations 
\citep{2006ESASP.604..183K} 
show that the accreting mechanism is most likely Wind Roche Lobe Overflow (WRLOF). 
However, \cite{2010ApJ...723.1188S} as well as \cite{2025NewA..12102452Z}
found that the accretion rate is considerably below the expected value. 
To shed light on this discrepancy, we performed new observations 
of the flickering during the 2025 minimum.  
These allow us to calculate the temperature and radius of the flickering source,
accretion disc luminosity, and estimate the mass of Mira~B.

\section{Observations}
\label{s.obs}

The observations were secured with three telescopes: 
{\bf (i)} the 50/70 cm Schmidt telescope  
\citep{1982BlDok..35..729G, 1987BlDok..40....9T},
{\bf (ii)} the 1.5m  AZ1500 telescope  \citep{2025arXiv250818752S}
of the Rozhen National Astronomical Observatory, Bulgaria,
and {\bf (iii)}  the 40~cm  telescope 
of the Shumen University   \citep{2020BlgAJ..32..113K}.
The three  telescopes are each equipped with CCD camera and rotating filter wheel.  
On the 50/70 cm Schmidt telescope 
we used full the CCD frame which covers a field of view 
$71 \times 71$~arcmin.

Comparison stars were  
BD$-03$~356 (B=11.140, V=10.645)
and  
HD~14411 (B=10.779, V=9.347). 
For control of the data processing we also used 
HD~14627 (B=9.456,  V=9.004),  
HD~14223 (B=10.055, V=9.491), 
and 
BD$-03$~364 (B=10.538, V=9.931). 
The magnitudes are taken from 
\cite{2022yCat.1360....0G} and 
from \cite{2014AJ....148...81M}.
The resulting light curves span from 3.7 to 4.9~hr.
Table~\ref{tab2} gives full details of each run including light curve statistics. 


In the optical V band  Mira pulsates between $2.5 < m_V < 9.0$ magnitude   
\citep{1997JAVSO..25..115H, 2009AcA....59..169G}. 
As the flickering variability is related to the companion, 
it is essential to observe at a time when
Mira~A is faint, and our new data are when $m_V \ge  8.3$. 

\begin{figure*}
\includegraphics[trim= 0.0cm 3.5cm 0.0cm 0.0cm, width=0.9\textwidth, angle=0]{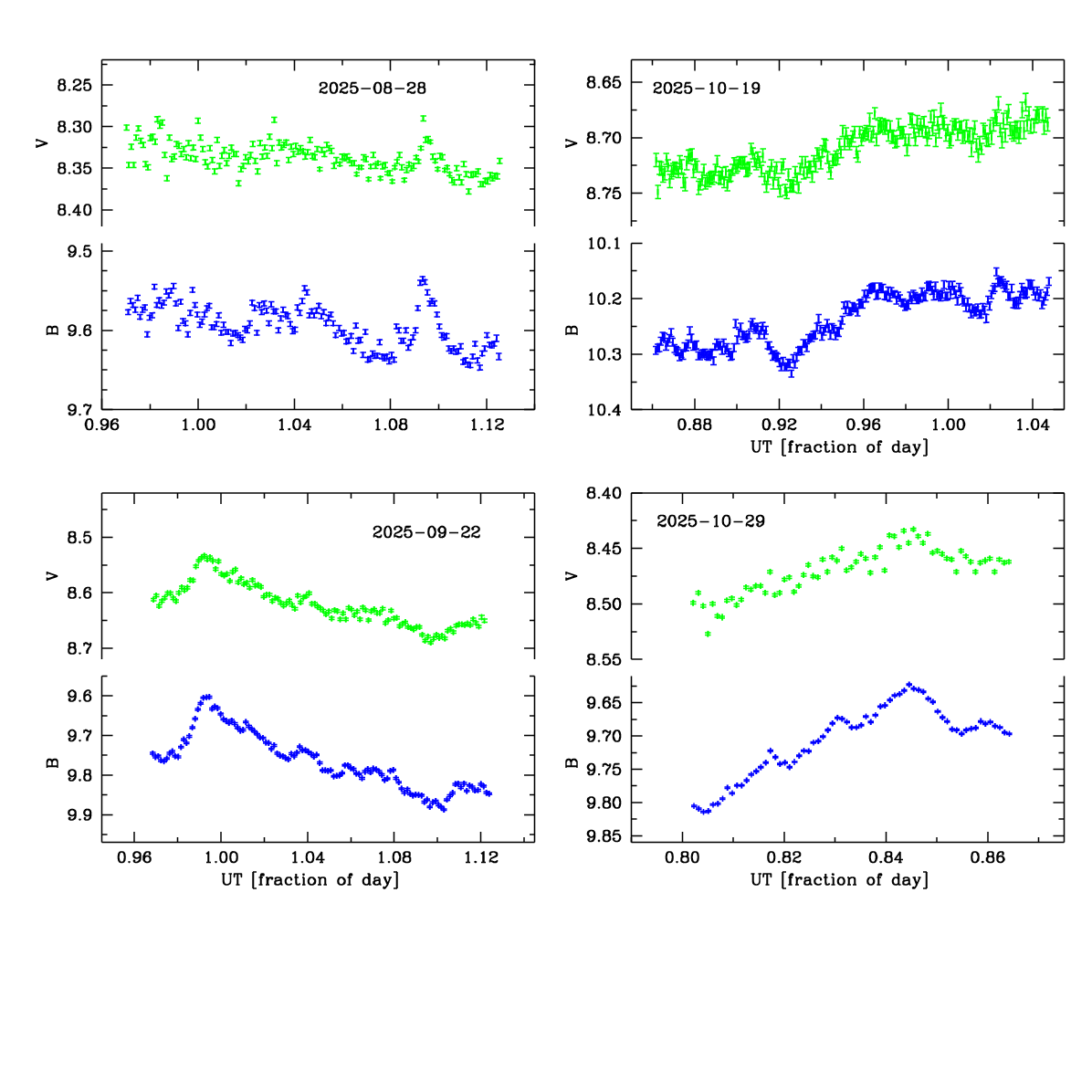}
 \caption[]{Short term variability of Mira in $B$ and $V$ bands. An intra-night 
           variability with an amplitude $\Delta B \sim 0.15$ mag is visible. 
           The date of observations is marked on each panel.}
\label{f.fli}	 
\end{figure*}

Part of our observations of omicron Ceti are plotted in Fig.~\ref{f.fli}. 
The intra-night variability (flickering) is evident on all nights. 
The peak-to-peak amplitude in  V band is 0.09 --0.16~mag,  
and  in  $B$ band it is 0.11--0.28~mag.
The variability in the two bands is synchronized, 
however the amplitude is larger in B band, 
similar to the symbiotic recurrent novae T~CrB and RS~Oph.

\section{Results}
\label{s.res}

\begin{figure*}
\includegraphics[trim= 0.0cm 1.0cm 0.0cm 8.0cm, width=0.9\textwidth, angle=0]{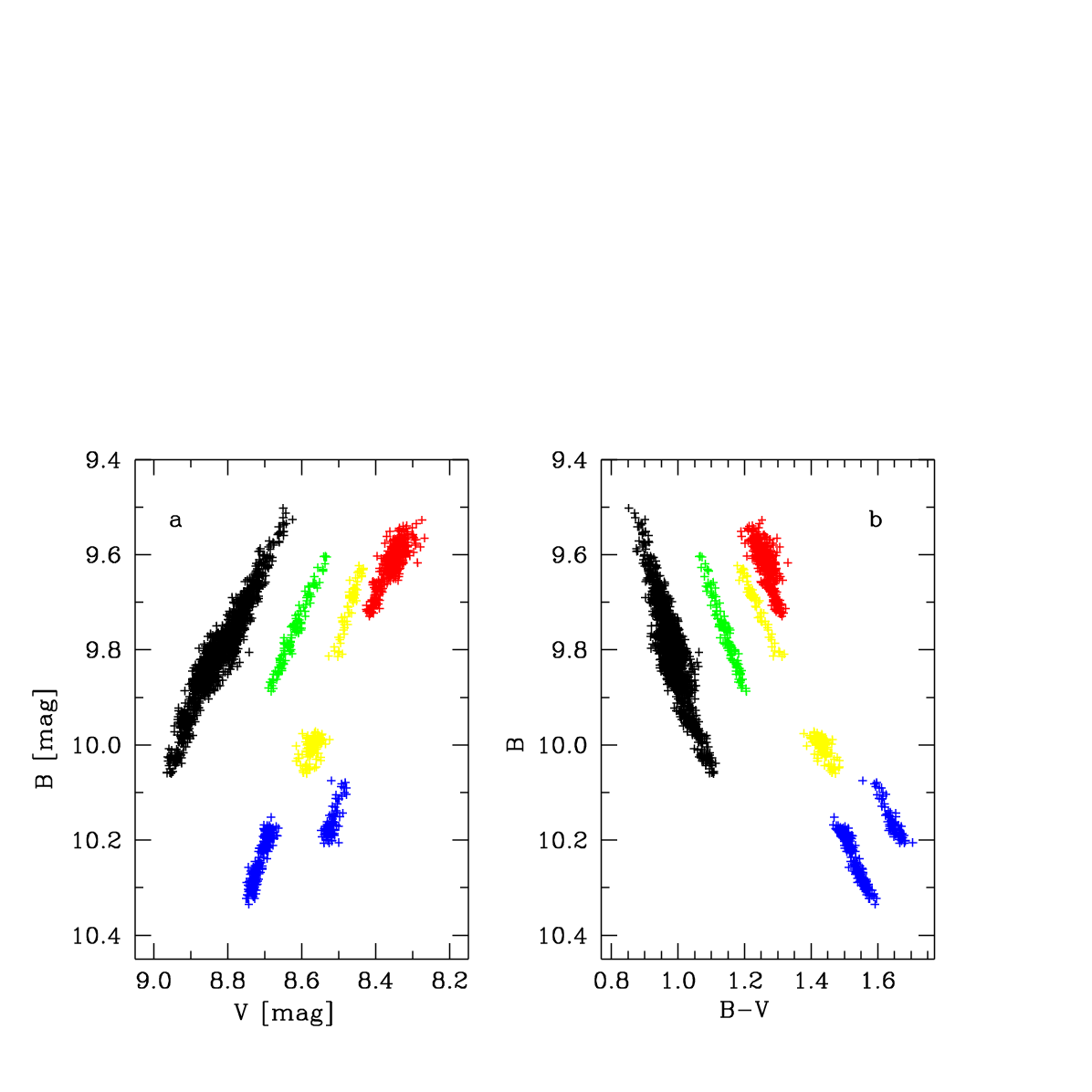}
  \caption[]{ {\bf a)} B-band  versus V-band magnitude. 
              {\bf b)} Colour magnitude diagram. The colours are as follows: 
data from 2024-11 (black),  2025-08-28 and 2025-08-29 (red), 2025-09-22 (green), 2025-10-19 and 2025-10-23 (blue),
2025-10-29 and 2025-10-31 (yellow). }
\label{f.BV}	 
\end{figure*}

In Fig.~\ref{f.BV}a we plot the  $B$ versus  $V$ magnitude. 
Fig.~\ref{f.BV}b represents the colour-magnitude diagram -- $B$ magnitude 
versus  $B-V$ colour. For each night the star becomes redder when it gets fainter. 
The positions of data on this figure 
are related to the brightness variations during the Mira cycle, 
as well as the variable luminosity of accretion disc (see Sect.~3.2).

\begin{table*}                                                                
  \caption{Photometry of Mira.  In the table are given: date of observation (in format YYYY-MM-DD),  
           UT-start and UT-end of the run (in format HH:MM), band, 
	   number of the data points, 
	   minimum, maximum, average and median magnitudes in the corresponding band, 
	   standard deviation of the mean, 
	   amplitude,  
	   typical observational error.  }
  \begin{tabular}{l c c  c | r  r r  r r r r r r} 
            &	  &	   &	            &	     &         &	 &	   &	   &	   &	   \\
 date	    & UT  &   band &   $N_{pts}$    &  min   &  max    & average & median  & stdev & ampl. & merr  \\
telescope   &	  &        &                & [mag]  & [mag]   &  [mag]  & [mag]   & [mag] & [mag] & [mag] \\
 \hline
	    &	           &	 &	    &	     &         &	 &	   &	   &	   &	   \\
 2025-08-28 & 23:17-03:01  & $B$ & 156x5s   & 9.535  &  9.647  &  9.5942 &  9.5950 & 0.026 & 0.112 & 0.004 \\
 50/70cm    &	           & $V$ & 157x3s   & 8.290  &  8.378  &  8.3376 &  8.3380 & 0.018 & 0.088 & 0.003 \\
	    &	           &	 &	    &	     &         &	 &	   &	   &	   &	   \\ 
 2025-08-29 & 22:28-02:59  & $B$ & 186x5s   & 9.567  &  9.729  &  9.6446 &  9.6395 & 0.044 & 0.162 & 0.003 \\
 50/70cm    &	           & $V$ & 189x3s   & 8.326  &  8.425  &  8.3747 &  8.3720 & 0.026 & 0.099 & 0.003 \\ 
	    &	           &	 &	    &	     &         &	 &	   &	   &	   &	   \\ 
 2025-09-22 & 23:14-02:57  & $B$ & 120x15s  & 9.603  &  9.887  & 9.7645  &  9.772  & 0.071 & 0.284 & 0.002 \\ 
 50/70cm    &              & $V$ & 118x5s   & 8.533  &  8.690  & 8.6212  &  8.627  & 0.038 & 0.157 & 0.002 \\  
	    &	           &	 &	    &	     &         &	 &	   &	   &	   &	   \\		
 2025-10-19 & 20:40-01:08  & $B$ & 233x10s  & 10.152 & 10.335  & 10.2367 & 10.2240 & 0.047 & 0.183 & 0.008 \\
 40cm	    &              & $V$ & 232x4s   & 8.668  &  8.750  &  8.7102 &  8.7065 & 0.020 & 0.082 & 0.009 \\  
 	    &	           &	 &	    &	     &         &	 &	   &	   &	   &	   \\
 2025-10-23 & 21:14-00:02  & $B$ & 78x10s   & 10.075 & 10.206  & 10.1611 & 10.1710 & 0.035 & 0.131 & 0.008 \\
 40cm	    &              & $V$ & 78x4s    &  8.479 & 8.547   & 8.5173  & 8.5185  & 0.016 & 0.068 & 0.008 \\	
 	    &	           &	 &	    &	     &         &	 &	   &	   &	   &	   \\
 2025-10-29 & 19:14-20:45  & $B$ & 67x5s    & 9.623  & 9.814   & 9.7066  & 9.6910  & 0.052 & 0.191 & 0.002 \\
  1.5m	    &		   & $V$ & 67x1s    & 8.433  & 8.527   & 8.4699  & 8.4670  & 0.021 & 0.094 & 0.002 \\ 
 	    &	           &	 &	    &	     &         &	 &	   &	   &	   &	   \\
 2025-10-31 & 22:00-23:32  & $B$ & 110x5s   & 9.972  & 10.060  & 10.0075 & 10.0015 & 0.023 & 0.088 & 0.003 \\
 1.5m	    &              & $V$ & 110x0.5s & 8.536  & 8.609   & 8.5707  & 8.5720  & 0.015 & 0.073 & 0.003 \\
 	    &	           &	 &	    &	     &         &	 &	   &	   &	   &	   \\
  \label{tab2}
  \end{tabular}
\end{table*}

The distance to Mira is estimated to be $92 \pm 10$ pc from the $Hipparcos$ parallax ($10.91\pm 1.22$~arcsec) 
\citep{2007A&A...474..653V}. 
Other values using different relations include: 
$105 \pm 7$ pc   \citep{1989MNRAS.241..375F},  
$107 \pm 12$ pc  \citep{2003A&A...403..993K},  
$115 \pm 7$ pc   \citep{2008MNRAS.386..313W}.  
We adopt a distance of 100 pc.  
The interstellar extinction at this distance is less than 0.01 mag \citep{2019A&A...625A.135L}. 
In such wide binary systems, the white dwarf is located outside of the dust cocoon associated with the Mira star \citep{2009AcA....59..169G} 
and we hence assume no extinction for Mira~B.

\subsection{Flickering}
For analysis of the flickering,
\citet{1992A&A...266..237B}    
suggests 
the light curve of the  intra-night variability 
to be separated into two parts -- constant light and variable (flickering) source. 
Following this procedure,    
we calculate the flux of the flickering light source 
as $F_{fl1}=F_{av}-F_{min}$, where $F_{av}$ is the average flux 
during the run and $F_{min}$ is the minimum flux during the run
(corrected for the typical error of the observations).
A slightly different method is proposed by 
\citet{2011ApJ...737....7N}.   
They used for the flickering source $F_{fl2}=F_{max}-F_{min}$, where $F_{ max}$ 
is the maximum flux during the run.  
In fact, the method of 
\citet{1992A&A...266..237B}                 
evaluates the average brightness of the flickering source, 
while that of 
\citet{2011ApJ...737....7N}               
 its maximal brightness. 
More details can be found in Sect.4 of \citet{1992A&A...266..237B}
and Sect.6.5 of \cite{2011ApJ...737....7N}).  
$F_{fl1}$ and $F_{fl2}$  have been calculated for each band, using the values 
given in Table~\ref{tab2}.  
To convert the observed magnitudes in fluxes, we use the calibration 
for a zero magnitude star 
$F_0 (B) = 6.13268 \times 10^{-9}$  erg cm$^{-2}$ s$^{-1}$ \AA$^{-1}$,     $\lambda_{eff}(B)=4371.07$~\AA, 
$F_0 (V) = 3.62708 \times 10^{-9}$  erg cm$^{-2}$ s$^{-1}$ \AA$^{-1}$ and  $\lambda_{eff}(V)=5477.70$~\AA\ 
as given in the Spanish virtual observatory 
Filter Profile Service  
\citep{2020sea..confE.182R}.  
To calculate the temperature, we use  the calibration for the $(B-V)$ colour 
of a black body (Table 18 in \citealt{1992msp..book.....S}). 
Using the temperature, distance $d=100$~pc, $F_{fl1}$,  and $F_{fl1}$, 
we calculate the radius and the luminosity of the flickering source.

In Table~\ref{t.2} are given: the colours 
$(B-V)_{01}$ and  $(B-V)_{02}$ of the flickering source,  
T$_{fl1}$  and  T$_{fl2}$ (temperature of the flickering source), 
R$_{fl1}$  and  R$_{fl2}$ (radius the flickering source), 
L$_{fl1}$  and  L$_{fl2}$ (luminosity the flickering source).
$(B-V)_{01}$, T$_{fl1}$,  R$_{fl1}$ and L$_{fl1}$  are calculated  using the average flux,  
following \cite{1992A&A...266..237B},  
$(B-V)_{02}$, T$_{fl2}$, R$_{fl2}$ and  L$_{fl2}$ --  using the maximum flux,  
following  \cite{2011ApJ...737....7N}. 

Using the method of \citet{1992A&A...266..237B},    
we find  for our runs:  
$ (B-V)_{01}$ in the range from 0.55 to 1.1,
corresponding to temperature in the range $4200-6200$~K,
radius of the flickering source $0.30 - 0.85$~$R_\odot$, 
and luminosity  $0.1-0.2$~L$_\odot$.  
Using the method of \citet{2011ApJ...737....7N}, 
we find  similar results for the colour and temperature:  
$ (B-V)_{02}$ in the range from 0.5 to 1.2,    
corresponding to temperature in the range $3800-6600$~K.  
The method of 
\citet{2011ApJ...737....7N}    
gives larger values
for the radius ($0.4-1.5$~$R_\odot$)  
and for the luminosity ($0.22-0.46$~L$_\odot$), 
because it refers to the maximal brightness.

\subsection{Optical luminosity of Mira~B}

The variability generated in accretion discs 
produces light curves that are phenomenologically similar across 
active galactic nuclei, X-ray binaries, and accreting white dwarfs. 
There is an amplitude-flux relation, which is  a result 
of the universality of the accretion physics from proto-stars still
in the star-forming process to the supermassive black holes at the
centers of galaxies  \citep[e.g.][and references therein]{2005MNRAS.359..345U, 
 2015SciA....1E0686S, 2016A&A...593L..17K}. 
This relation is probably connected 
 with the viscosity \citep{2004MNRAS.348..111K}.
To estimate the luminosity of the accretion disc around the white dwarf, 
we use the observed relationship between the 
amplitude of the flickering ($\Delta F$)
and the average flux $F_{av}$. 
For accreting white dwarfs these two parameters are connected  as  
$\Delta F / F_{av} = 0.362 \pm 0.045$   
 \citep{2016MNRAS.457L..10Z}.  
From this  amplitude-flux  relation,  we adopt 
\begin{equation}  
2 L_{fl1} = 0.362 \, L_{d}, 
\label{eq.L1}
\end{equation} 
\begin{equation}  
L_{fl2} = 0.362 \, L_{d}, 
\label{eq.L2}
\end{equation} 
where  $L_{d}$ is the accretion disc luminosity.
In Table~\ref{t.3} is given the calculated $L_{d}$ 
for the  past flickering observations \citep{2019BlgAJ..31..110Z, 2025NewA..12102452Z}, as well as the new data. 
The calculated luminosity of the  accretion disc  is in the range 
from 0.4 to 1.3~L$_\odot$, with average 
$L_{d} =  0.91 \pm 0.28$~L$_\odot$.


\begin{table*}
  \caption{Flickering source of Mira~B.  
$(B-V)_{01}$, T$_{1}$, R$_{fl1}$, and  $L_{fl1}$  
       are dereddened colour, 
       temperature, radius and luminosity  of the  flickering source calculated following Bruch (1992), 
$(B-V)_{02}$, T$_2$, R$_{fl2}$ and  L$_{fl2}$ --  following Nelson et al. (2011), 
       see Sect.~\ref{s.res} for details. }
  \begin{tabular}{l | c  c  c  c | c  c c c c c}
 & 	     &       &  	   &		   & 	     &        & \\
 date	    & $(B-V)_{01}$ & T$_{1}$  &  $R_{fl1}$  &  $L_{fl1}$ &  $(B-V)_{02}$ & T$_{2}$ &  R$_{fl2}$ & L$_{fl2}$ & \\ 
 	    &           & [K] & $[R_\odot]$ & [L$_\odot$]  &               & [K] & $[R_\odot]$ & [L$_\odot$] & \\
 \hline
	    & 	     &       &  	   &		   & 	     &        & \\
 2025-08-28 &   0.9721  &  4439  &    0.7671  & 0.205  &  0.9950  &  4375  &	1.1960  &   0.470 & \\ 
 2025-08-29 &   0.7248  &  5342  &    0.5024  & 0.184  &  0.7372  &  5285  &	0.7328  &   0.376 & \\
 2025-09-22 &   0.5434  &  6229  &    0.3656  & 0.180  &  0.4878  &  6576  &	0.5028  &   0.424 & \\
 2025-10-19 &   0.5738  &  6039  &    0.2880  & 0.099  &  0.6619  &  5628  &	0.4895  &   0.216 & \\
 2025-10-23 &   0.9606	&  4471	 &    0.5315  &	0.101  &  0.818   &  4941  &	0.6525  &   0.228 & \\
 2025-10-29 &   0.5756	&  6028	 &    0.3854  &	0.176  &  0.468   &  6702  &	0.3978  &   0.286 & \\
 2025-10-31 &   1.1014	&  4079	 &    0.8773  &	0.191  &  1.220   &  3770  &	1.5944  &   0.460 & \\
 \label{t.2}
 \end{tabular}
 \end{table*}

\begin{table}
  \caption{Accretion disc luminosity, $L_{d}$ of  Mira~B. 
  The first column gives date of observation.
  The second and the third columns give the estimated $L_d$
  using Eq.\ref{eq.L1} and Eq.\ref{eq.L2}, respectively (see Sect.~\ref{s.res} for details). }
 \begin{tabular}{l | c  c  c c c}
 & 	     &           &         & \\
 date	     &   $L_{d}$ & $L_{d}$ & \\ 
             & [$L_\odot$] & [$L_\odot$] & \\
 	     & from  Eq.\ref{eq.L1}  &  from  Eq.\ref{eq.L2}       & \\
 \hline
             &         &          &  \\ 
2018-09-07   &  0.345  &   0.584  &  \\
2018-09-09   &  0.439  &   0.977  &  \\
2024-11-24   &  1.163  &   1.165  &  \\
2024-11-25   &  0.787  &   0.861  &  \\
2024-11-26   &  1.012  &   0.992  &  \\
2024-11-27   &  0.983  &   1.358  &  \\
2025-08-28   &  1.132  &   1.298  &  \\
2025-08-29   &  1.018  &   1.038  &  \\
2025-09-22   &  0.997  &   1.172  &  \\
2025-10-19   &  0.547  &   0.595  &  \\	   	
2025-10-23   &  0.559  &   0.629  &  \\
2025-10-29   &  0.972  &   0.791  &  \\
2025-10-31   &  1.056  &   1.272  &  \\  
             &         &          &  \\
average      & \multicolumn{2}{c}{$0.91\pm0.28$}   &  \\
 \label{t.3}
 \end{tabular}
 \end{table}

\subsection{Accretion luminosity of Mira~B}
\label{s.L}

The  luminosity of the accretion disc of an accreting white dwarf depends 
on the mass of the white dwarf, its radius, and the mass accretion rate: 
\begin{equation}
L_{d} =  \frac{G \;  M_{wd} \;  \dot M_a}{2 \; R_{wd}},  \\
\label{eq.Ld}
\end{equation}
where $\dot M_a$ is the mass accretion rate, 
$M_{wd}$ is the mass of the white dwarf, 
$R_{wd}$ is the radius of the white dwarf. 
For a standard accretion disc,  the disc luminosity 
is half of the total accretion luminosity.
The other half is emitted  by the boundary layer between 
the  accretion disc and the white dwarf.

A characteristic feature  of the white dwarfs is 
that the less massive white dwarfs have larger radii. 
For the radius of the white dwarf, 
we use the formula by P.~Eggleton  as given in 
\citet{1988ApJ...332..193V}:             
\\
\begin{equation}
\begin{aligned}
\frac{ R_{wd}}{ R_\odot } = 0.0114 
\left[  \left( \frac{M_{wd}}{M_{Ch}} \right)^{-2/3} - \left( \frac{M_{wd}}{M_{Ch}} \right)^{2/3} \right]^{1/2}  \\ 
\times \left[  1 + 3.5 \left( \frac{M_{wd}}{M_p} \right)^{-2/3} + \left( \frac{M_{wd}}{M_p} \right)^{-1} \right]^{-2/3}, \\
\\
\end{aligned}
\label{eq.Rwd}
\end{equation}
where $M_p $ is a constant $M_p= 0.00057$~M$_\odot$, 
$M_{Ch}=1.44$~M$_\odot$ is the Chandrasekhar mass limit for a white dwarf. 
We note in passing that the observed masses and radii of 
white dwarfs agree with this formula
\citep[e.g.][and references therein]{2025MNRAS.tmp.1726B}.
This mass-radius relation gives radius $R_{wd} = 8614$~km for a $M_{wd} = 0.6$~M$_\odot$, 
$R_{wd} = 10675$~km for a $M_{wd} = 0.4$~M$_\odot$, 
$R_{wd} = 14075$~km for a $M_{wd} = 0.2$~M$_\odot$. 


\subsection{Bondi-Hoyle-Lyttleton accretion}
\label{s.BHL}

The orbital period of Mira~A+B binary is $\ge 500$~yr.
In such wide binaries the white dwarf captures a fraction of the 
wind of the red giant. In standard Bondi-Hoyle-Lyttleton wind accretion 
(for a review, see \cite{2004NewAR..48..843E}),
the accretion rate onto the white dwarf depends on the orbital and stellar parameters
\citep{1939PCPS...35..405H, 1944MNRAS.104..273B}: 
\begin{equation}
  \dot M_{acc} = \frac {R_a^2}{ 4 \; a^2} \; \dot M_w \; , 
\label{eq.Ma}
\end{equation}
where $\dot M_w$ is the wind mass loss rate of the red giant (Mira~A),
$a$ is the semi-major axis of the binary orbit,  
$R_a$ is the accretion radius of the white dwarf:
\begin{equation}
  R_a =  \frac{ 2 \;  G \; M_{wd}}{v_w^2 + v_{orb}^2 + c_s^2} \; ,
\label{eq.Ra}
\end{equation}
where $v_w$ is the wind velocity, 
$v_{orb}$ is the orbital velocity,
$c_s$ is the speed of sound in the wind at distance $a$. 
We adopt $c_s=1$~km~s$^{-1}$ \citep{2020MNRAS.499.1531S}. 
The semi-major axis is related to the orbital  period 
by the Kepler's third law:
\begin{equation}
  P_{orb}^2 = \frac{4 \; \pi^2 \; a^3}{ G \; (M_1 + M_{wd}) } \; ,
\label{eq.K3}
\end{equation}
where $M_1$ is the mass of the red giant primary (Mira~A). 
The orbital velocity of the white dwarf $v_{orb}$ is 
\begin{equation}
  v_{orb} = \frac{2 \, \pi \;  M_1 \; a} {(M_1 + M_{wd}) \; P_{orb}} .
\label{eq.Vorb}
\end{equation}
For the mass loss rate of Mira~A we adopt 
$M_w = 4.4\times 10^{-7}$~M$_\odot$~yr$^{-1}$ and 
$v_w = 6.7$~km~s$^{-1}$ 
\citep{1998ApJS..117..209K}.  
Omicron Ceti is a very wide binary system.
\citet{2018MNRAS.477.4200S} collected data for Mira~B positions 
relative to Mira~A and fitted them with a circular orbit inclined at
$i = 67^0$, with period 945 yr and angular separation 1.03 arcsec.
For a distance $d =100$~pc, it gives 
a physical separation between Mira A and B of $a = 103$~au.

Using Eq.~\ref{eq.K3} and the observed parameters P$_{orb}$ and $a$ we estimate the mass of the system. 
We calculate the orbital velocity V$_{orb}$ via Eq.~\ref{eq.Vorb}
for a range of white dwarf masses (from $0.1$~M$_\odot$ to $1.43$~M$_\odot$). 
We use Eq.~\ref{eq.Ma} and Eq.~\ref{eq.Ra}
to estimate the mass acretion rate $\dot M_{acc}$. 
The  radius of the white dwarf for each mass is calculated via Eq.~\ref{eq.Rwd}
and then the accretion luminosity L$_d$ is obtained from Eq.~\ref{eq.Ld}. 
We compare the resulting values for L$_d$ with the value derived from our observations 
and achieve agreement at $M_{wd} = 0.42 \pm 0.02$~M$_\odot$  and $M_1 \approx 0.8$~M$_\odot$. 
Such a white dwarf will accrete at a rate 
$\dot M_a \approx 2.1 \times 10^{-9}$~M$_\odot$~yr$^{-1}$ 
and will emit $L_d \approx 0.93$~L$_\odot$.

\subsection{Wind Roche-Lobe Overflow  (WRLOF) }
\label{s.WRLOF}

Wind Roche-Lobe Overflow (WRLOF) is  proposed 
by  \cite{2007ASPC..372..397M}  
for symbiotic binaries. 
This is a mass-transfer mechanism where the stellar wind 
(particularly if it is dense and slow),
fills the Roche-lobe of the red giant 
and is transferred to the hot component through the inner Lagrangian point.
This mechanism is a hybrid between the normal Roche-lobe overflow  
and standard wind accretion (Bondi-Hoyle-Lyttleton accretion). 
Numerical simulations \citep{2012BaltA..21...88M, 2017MNRAS.468.3408D}  
have shown that in case of 
WRLOF the subsequent mass-transfer rate is
at least an order of magnitude greater than the analogous Bondi-Hoyle-Lyttleton value.
Here we will assume that it is 10 -- 20 times greater.

In agreement with WRLOF scenario, 
the high resolution {\sc Chandra} and {\sc Hubble Space telescope}
observations have shown that in the mass exchange between Mira~A and Mira~B 
in addition to wind accretion there is evidence for Roche-lobe like overflow
\citep{2005ApJ...623L.137K, 2006ESASP.604..183K}. 

Assuming that in case of WRLOF the 
mass accretion rate is  10 times higher than the standard Bondi-Hoyle-Lyttleton value,
and using the above equations we find that a white dwarf having accretion disc luminosity
$L_d \approx 0.91$~L$_\odot$ accreting through WRLOF, has  
to be a low mass,  with $M_{wd} = 0.21 \pm 0.02$.
This value of the mass is the lower limit because in Eq.~\ref{eq.Ld} 
the inclination of the accretion disc is not taken into account. 
If we adopt  $L_d = 0.5 \, G \,  M_{wd} \, \dot M_a \, \cos i \, R_{wd}^{-1} \,$ $\;$ and
$i= 67^0$, 
then we get $M_{wd} = 0.28$~M$_\odot$ for a WRLOF with 10 times higher accretion rate than Bondi-Hoyle-Lyttleton accretion and $M_{wd} = 0.23$~M$_\odot$ for a WRLOF with a 20 times higher accretion rate. 

The WRLOF mechanism suggests that Mira B is a low mass white dwarf 
with mass $0.24\pm 0.04$~M$_\odot$ accreting at a rate of $6.8 \times 10^{-9}$~M$_\odot$ yr$^{-1}$. 
In other words the white dwarf captures $\approx 1.5\%$ of the wind of the giant.

\section{Discussion}

Here we explore how varying various critical parameters affects our overall conclusions. 
Symbiotic Miras in general have very long orbital periods, 
longer than 100~yr \citep{2013ApJ...770...28H}. 
In Sect~\ref{s.res}, we used $P_{orb} = 945$ yr 
\citep{2018MNRAS.477.4200S}.  
Earlier estimates give shorter periods: 498 yr \citep{2002ApJS..139..249P} 
and  610 yr \citep{2007ApJ...662..651I}. 
If we assume $P_{orb} = 610$ yr, than we get an almost identical value of
$M_{wd} = 0.24$~M$_\odot$, because this will change $v_{orb}$. 
However the main factor in  Eq.~\ref{eq.Ra} is $v_w$.

In Sect~\ref{s.res}, we used speed of sound $c_s=1.0$~km~s$^{-1}$.
If we assume a two times larger value  $c_s=2.0$~km~s$^{-1}$ or two times smaller value $c_s=0.5$~km~s$^{-1}$ the result will be 
practically the same because 
the motion of the accreting object is supersonic and 
the $c_s$ is a minor factor in Eq.~\ref{eq.Ra}. 

If we use two times lower mass loss rate of Mira~A, 
$M_w = 2.2 \times 10^{-7}$~M$_\odot$~yr$^{-1}$, and 
$v_w = 6.7$~km~s$^{-1}$,   
we get the mass of the white dwarf $\approx 0.27$~$M_\odot$.
If we use different values for the mass loss of Mira~A, 
$M_w = 3.6 \times 10^{-7}$~M$_\odot$~yr$^{-1}$ and 
$v_w = 4.8$~km~s$^{-1}$ as given by \cite{2005A&A...439..171H}, 
we   derive an even lower mass of the white dwarf $\approx 0.17$~$M_\odot$.

Significant changes in the accretion rate have been inferred in the past for Mira B 
\citep{2006ApJ...649..410W}.  
We also detect variations in the disc luminosity  in the range
$0.3 \le L_d \le 1.4$~L$_\odot$ (see Table \ref{t.3}). 
The amplitude of the flickering in B band
in our observations is $0.1-0.28$~mag, which is 
similar to the amplitudes $0.15-0.30$~mag observed by \citet{2010ApJ...723.1188S}.

In wide binaries as Mira~A+B,  the white dwarf  accretes  
material via gravitational capture of the red-giant wind. 
\citet{2010ApJ...723.1188S}  
estimated  $\dot M_a  \sim 10^{-10}$~M$_\odot$~yr$^{-1}$ assuming 
$M_{wd}=0.6$~$M_\odot$. 
 \citet{2018MNRAS.477.4200S} also assume  $M_{wd} = 0.6$~$M_\odot$.
This assumption was first made by \citet{1985ApJ...297..275R}
on the basis that two thirds of the DA white dwarfs have masses 
in a narrow range $0.58 \pm 0.10$~$M_\odot$ \citep{1979A&A....76..262K}.
Our estimate indicates that in fact the white dwarf companion
(Mira~B) is a lower mass white dwarf $M_{wd} \approx 0.42$~$M_\odot$ 
(in the case of  Bondi-Hoyle-Lyttleton accretion)
and a more likely value $M_{wd} \approx 0.24$~$M_\odot$ 
(in the case of WRLOF). 
The result explains  some early findings
that the mass accretion rate is below the theretical value for  WRLOF
and even  for Bondi-Hoyle-Lyttleton accretion (see Sect. 5 in \citet{2010ApJ...723.1188S}).
We consider the value obtained suggesting WRLOF is more probable,
because the high resolution observations have shown that in the mass exchange between
Mira~A and Mira~B there 
is evidence 
for Roche-lobe like overflow \citep{2005ApJ...623L.137K, 2006ESASP.604..183K}.
We note that \citet{1984ApJ...287..785J}
suggested that the low X-ray emission of the system can be explained 
if the secondary component is either a main-sequence red dwarf or an extremely low mass white dwarf.
Our result is in agreement with the suggestion for a low mass white dwarf.

Having in mind our favoured result $M_{wd}=0.24$~$M_\odot$
and using Eq.~\ref{eq.K3}, we estimate mass of Mira~A,  $M_1 \approx 1$~~$M_\odot$.
This value is highly sensitive to the adopted orbital period.
If we assume $P_{orb} = 610$~yr, we find $M_1 \approx 2.7$~~$M_\odot$.
Nevertheless, both values are in the range of 
masses of the stars on the asymptotic giant branch  0.5-4~$M_\odot$ \citep{1993ApJ...413..641V}. 

\citet{2005ApJS..156...47L}
studied a sample of 348 white dwarfs from the Palomar Green Survey and obtained a mass distribution showing a main peak centered near 0.6~M$_\odot$, a low-mass component centered near 0.4~M$_\odot$,
and a high-mass component above about 0.8~M$_\odot$. 
Our result for the mass of Mira B ($M_{wd} \approx 0.42$~$M_\odot$, assuming Bondi-Hoyle-Lyttleton accretion) is close to the low-mass component peak of the Palomar Green Survey distribution. 
If we assume WRLOF accretion, we obtain a value of 0.24~M$_\odot$ 
for the mass of the white dwarf, which is below 
the lowest mass (0.32~M$_\odot$) observed in the Palomar Green  white dwarfs. 
More resent surveys, have shown that white dwarfs with mass less than 0.3~M$_\odot$ do exist.
In the Gaia~DR2 there are 5762 extremely low mass white dwarf candidates with masses 
below 0.3~M$_\odot$ \citep{2019MNRAS.488.2892P}. 
\citet{2007ApJ...660.1451K} reported that there are four white dwarfs from the SDSS 
with masses even below 0.2~M$_\odot$. 

The numerical simulations, high resolution observations 
as well as our results indicate that
Mira~B is most likely part of the group of extremely low mass white dwarfs.

\section{Conclusions}

We report 22.3 hours of simultaneous observations in B and V bands 
of the flickering of Mira obtained with the telescopes 
of National Astronomical Observatory Rozhen and 
Shumen University, Bulgaria. 
The observations were performed during August-October 2025. 
The amplitude of the flickering was 0.11-0.28 mag in $B$ band. 
We find that the average luminosity of the accretion disc around Mira~B is 
$0.91 \pm 0.28$~$L_\odot$.  
Combining the equations for disc luminosity, mass accretion rate, 
and mass-radius relation for white dwarfs,
we estimate that the white dwarf companion
(Mira~B) is a low mass white dwarf with $M_{wd} \approx 0.42 \pm 0.04$~$M_\odot$ 
in the case of 
Bondi-Hoyle-Lyttleton accretion
or an extremely low mass white dwarf with $M_{wd} \approx 0.24 \pm 0.04$~$M_\odot$
in the case of WRLOF.

\section{Acknowledgments}
This work is part of the project KP-06-H98/8 "Accretion flows in binary stars"
(Bulgarian National Science fund).   
The research infrastructure is supported by 
the National Roadmap for Research Infrastructure coordinated 
by the Ministry of Education and Science of Bulgaria. 
DM, BB and GY acknowledge  Scientific Research Fund of Shumen University. 
We thank an anonymous referee for making very valuable suggestions.

\section*{Data Availability} 
The data are available for download at zenodo.org/records/18756532

\bibliographystyle{mnras}
\bibliography{ref5.bib}

\bsp	
\label{lastpage}
\end{document}